# Multipath Interference Analysis of IR-UWB Systems in Indoor Office LOS Environment

Luo Chao
School of Electronics and Information Engineering
Harbin Institute of Technology
Harbin, China
chluochao@163.com

Wu Xuanli
School of Computer Science and Technology
School of Electronics and Information Engineering
Harbin Institute of Technology
Harbin, China
xlwu2002@hit.edu.cn

Cao Yang
School of Electronics and Information Engineering
Harbin Institute of Technology
Harbin, China
xiaoyang_cao@126.com

*Abstract*—Bit error rate (BER) performance of impulse radio Ultra-Wideband (UWB) systems in the presence of intra-symbol interference, inter-symbol interference, multiuser interference and addictive white Gaussian noise (AWGN) is presented in this paper. By analyzing the indoor office LOS channel model defined by IEEE 802.15.4a Task Group and deducing the variance for intra-symbol interference (IASI), inter-symbol interference (ISI) and multiuser interference (MUI), the system BER expression is obtained and verified by MATLAB simulations. Through comparing the simulation results with and without intra-symbol interference, the conclusion that intra-symbol interference cannot be neglected is drawn-moreover, such interference will significantly decrease performance of UWB based wireless sensor networks (WSN). Then, the BER performance of UWB systems in multiuser environment is also analyzed and analysis results show that multiuser interference will further worsen the transmission performance of UWB systems.

## I. INTRODUCTION

As a high data rate, energy-efficient and low complexity wireless access technology, UWB is becoming an increasingly popular field for researchers. Many literatures indicate that UWB is one of the feasible technologies for wireless sensor networks [1]. And, the IEEE 802.15.4a Group Task choose UWB as one of the physical layer techniques [2]. However, in certain specific environments, especially indoor office environment, the dense multipath channel of UWB exhibits an obstacle to collect energy effectively with substantial multipath interference. There are two types of multipath interferences: One is caused by the interference of two adjacent data symbol (ISI). Another is caused by the interference between a pulse and its own multipaths, which is called intra-symbol interference (IASI). The available literatures in this area have already analyzed the system performance in the presence of narrowband interference, multiuser interference, and multipath interference [3-8]. Reference [3] and [4] mainly discussed multiuser interference and proposed the energy-detection receiver and non-linear filtering approach to mitigate such interference. Reference [5] mainly investigated the system performance of IR-UWB with narrow band interference (NBI) and concluded that NBI will greatly worsen system performance. Reference [6-8] have analyzed the inter-symbol interference (ISI) and figured out that ISI effect can be neglected when data transmission rate is low while for the high data rate, ISI can impose a significant influence on system performance. Nevertheless, the literatures above do not consider the intra-symbol interference (IASI) effect and do not present a persuasive proof that IASI can be neglected in IEEE 802.15.4a indoor office environment.

In this paper, we are going to prove that IASI cannot be neglected and establish a proper mathematical model to make a detailed research into two types of multipath interferences, i.e., intra-symbol interference and inter-symbol interference, as well as multiuser interference, and then analyse how they affect the BER performance of UWB system.

## II. UWB MULTIUSER AND MULTIPATH CHANNEL MODLE

In a TH-UWB system modulated by BPSK, the transmitted signal can be expressed as:

$$s^{(n,i)}(t) = \sum_{j=0}^{N_s-1} d^{(n,i)} \sqrt{E_p} p(t - jT_f - C_{i,j}^{(n)} T_c). \qquad (1)$$

where, $p(t)$ is a unit energy pulse waveform with energy $E_n$. $T_f$ is the mean pulse repetition period. $C_{i,j}^{(n)} = \{1,2,3,\ldots,N_h\}$ is the time-hopping sequence of the $i$-th bit of the $n$-th user, $T_c$ is the time-hopping slot time, $d^{(n,i)} \in \{-1,1\}$ represents the binary data sequence. One data symbol is conveyed with $N_s$ pulses.

The UWB channel model discussed in this paper is generated from the IEEE 802.15.4a indoor office LOS environment with path frequency dependence. In time domain, the impulse response of UWB system can be written as follows:

This work is supported by Specialized Research Fund for the Doctoral Program of Higher Education (New Teachers) (Grant No. 20092302120001), China Postdoctoral Science Foundation (Grant No. 20100471080), and Heilongjiang Postdoctoral Grant (Grant No. LBH-Z09153).

$$h(t) = f(t) * \sum_{l=1}^{L}\sum_{k=1}^{K} \alpha_{k,l} \delta(t - T_l - \tau_{k,l}). \tag{2}$$

And in frequency domain, it can be written as:

$$H(\omega) = F(\omega)\sum_{l=1}^{L}\sum_{k=1}^{K} \alpha_{k,l} \exp(-j\omega \cdot (T_l + \tau_{k,l})). \tag{3}$$

where, $L$ and $K$ denote the total number of clusters and rays. $\alpha_{k,l}$ is the tap weight of the $k$-th ray in the $l$-th cluster, and $\tau_{k,l}$ is the arrival time of $k$-th ray relative to $l$-th cluster arrival time $T_l$. And $F(\omega)$ denotes the frequency dependence of ray arrivals, which can be given by:

$$F(\omega) = C_0 (\omega / \omega_0)^{-\kappa}. \tag{4}$$

where, $C_0$ is a constant, $\kappa$ is the frequency dependence of the pathloss, and $\omega_0$ is the reference frequency. Furthermore, $F(\omega)$ can be expanded in Taylor series as

$$F(\omega) \approx F(\omega_c) + F'(\omega_c)(\omega - \omega_c). \tag{5}$$

where, $\omega_c$ is the center frequency. In indoor office LOS environment, $\kappa$ is considerably small and $F(\omega)$ is a slowly varying function of $\kappa$ within the applied frequency bandwidth [9]. Thus we can obtain the approximation of $F(\omega) \approx F(\omega_0)$ by ignoring the higher order terms of Taylor series.

It is noted that for IR-UWB systems using pulse based transmitter and receiver, the pulse has only positive and negative polarity. Thus, there is no need to consider random phase angle in equation (2).

Thus, the received signal can be demonstrated as:

$$r(t) = \sum_{n=1}^{N_u}\sum_{i=1}^{N_I} s^{(n,i)}(t - \tau^{(n)} + iT_f) * h(t) + n(t). \tag{6}$$

where, $n(t)$ denotes addictive white Gaussian noise and $\tau^{(n)}$ is the $n$-th user's reference delay relative to first user because of asynchronous transmission, assuming $\tau^{(1)} = 0$. $N_I$ represents the number of interfering pulses from the previous periods. $N_I = \lceil \tau_{\max}/T_f \rceil = \lceil \tau_{\max} R_b N_s \rceil$, $R_b$ is transmission bit rate, $\tau_{\max}$ is maximum multipath delay.

Without loss of generality, we suppose the first ray in the first cluster is the desired ray to be received, whose energy is $\Omega_0$. Then the delay of the $n$-th user resulting from different propagation distance and different transmitting time is $t_u^{(n)} = \tau^{(n)} - \tau^{(1)}$. Moreover, for the same user the delay of the $l$-th cluster relative to the first cluster is $t_c^{(n)} = T_l - T_1$. And in the same cluster, the delay of the $k$-th ray relative to the first ray is $t_p^{(k)} = \tau_{k,l} - \tau_{1,l}$. $\tau_{code}^{(n)} = (C_{i,j}^{(n)} - C_{0,j}^{(n)}) \cdot T_c$ denotes the TH code interval between the $i$-th interfering pulse and the current receiving pulse with $T_c$ corresponding to the hop width.

According to IEEE 802.15.4a channel model, the distribution of the cluster arrival times is given by a Poisson processes:

$$p(T_l / T_{l-1}) = \Lambda_l \exp[-\Lambda_l (T_l - T_{l-1})], l > 0. \tag{7}$$

where, $\Lambda_l$ is the cluster rate.

Similarly, the distribution of $\tau_{k,l}$ is modeled with a mixture of two Poisson processes as follows:

$$p(\tau_{k,l} / \tau_{(k-1),l}) = \beta\lambda_1 \exp[-\lambda_1(\tau_{k,l} - \tau_{(k-1),l})] \\ + (1-\beta)\lambda_2 \exp[-\lambda_2(\tau_{k,l} - \tau_{(k-1),l})], k > 0. \tag{8}$$

where, $\beta$ is the mixture probability, while $\lambda_1$ and $\lambda_2$ are the ray arrival rates.

According to probability theory, $t_c^{(l)}$ obeys Poisson distribution with two parameters $\Lambda$ and $l$, and the probability dense function (PDF) can be given by [9]

$$f_c(x) = \Lambda \exp(-\Lambda x)\frac{(\Lambda x)^{(l-2)}}{(l-2)!}. \tag{9}$$

Nevertheless, $t_p^{(k)}$ is described with a mixture Poisson processes and it is difficult to analyze its distribution function and probability dense function (PDF). Meanwhile, we notice that in indoor office LOS environment, $\beta$ is considerably small, which indicates that the occurrence of Poisson process with parameter $\lambda_1$ is very small and the Poisson process with parameter $\lambda_2$ is dominant. To simplify our computation, we take $t_p^{(k)}$ as a single Poisson process with parameter $\lambda_2$ and the PDF is given by

$$f_p(x) = \lambda_2 \exp(-\lambda_2 x)\frac{(\lambda_2 x)^{(k-2)}}{(k-2)!}. \tag{10}$$

The PDF of $\tau_{code(i)}^{(n)}$ follows

$$f_{code(i)}(x) = \begin{cases} 1/(2T_s) & x \in [-T_s, T_s] \\ 0 & elsewhere \end{cases}. \tag{11}$$

where, $T_s$ is the maximum time hopping position with $T_s \leq T_f$.

For the fading amplitude $\alpha_{k,l}$, it follows a Nakagami-m distribution with parameters ($\Omega, m$) according to [10]

$$pdf(\alpha_{k,l}) = \frac{2}{\Gamma(m_{k,l})}(\frac{m_{k,l}}{\Omega_{k,l}})^{m_{k,l}} \alpha_{k,l}^{2m_{k,l}-1} \exp(-\frac{m_{k,l}}{\Omega_{k,l}}\alpha_{k,l}^2). \tag{12}$$

where, $\Gamma(\cdot)$ corresponds to the Gamma function, $m$ is the Nakagami m-factor which is modeled as a lognormally distribution random variable, $E[\alpha_{k,l}^2] = \Omega_{k,l}$.

The mean power of different rays is expressed by

$$E[\alpha_{k,l}\alpha_{k_1,l_1}] = \begin{cases} \dfrac{\Omega_l \exp(-\tau_{k,l}/\gamma_l)}{\gamma_l[(1-\beta)\lambda_1 + \beta\lambda_2 + 1]} & k = k_1 \text{ and } l = l_1 \\ 0 & k \neq k_1 \text{ or } l \neq l_1 \end{cases}. \tag{13}$$

where, $\Omega_l$ corresponds to the integrated energy of the $l$-th cluster, and $\gamma_l$ is the intra-cluster decay time constant. $\gamma_l$ is linearly depended on the arrival time of the cluster,

$$\gamma_l \propto k_\gamma T_l + \gamma_0. \tag{14}$$

and the mean energy of the $l$-th cluster is given by

$$10\log(\Omega_l) = 10\log(\exp(-T_l/\Gamma)) + M_{cluster}. \quad (15)$$

### III. MULTIPTH INTERFERENCE MODLE

According to literatures [11-12], it is generally supposed that the interference between a pulse and its own multipaths can be ignored, namely the different multipath components coming from one pulse can usually be resolved and they do not result in intra-symbol interference (IASI). However, these literatures do not present a definite proof on this point of view. In the following part, we are going to prove that intra-symbol interference cannot be ignored in IEEE 802.15.4a indoor office LOS environment. According to the channel mode, we can obtain the mean ray interval and the mean cluster interval.

The mean ray interval ($\Delta_\tau$) is

$$E[\Delta_\tau] = \int_0^{+\infty} \Delta_p \lambda_2 e^{-\lambda_2 \Delta_p} d\Delta_p = \frac{1}{\lambda_2}. \quad (16)$$

And in indoor office LOS environment $\lambda_2 = 2.97(1/\text{ns})$ according to (16), the mean interval is approximately 0.34ns. However, the pulse duration mentioned in available literatures ranges from 0.5ns to 2ns. In this case, the mean ray interval is shorter than pulse duration and there is a probability that several adjacent rays may overlap with each other and result in intra-symbol interference.

Similarly, the mean cluster interval ($\Delta_c$) is

$$E[\Delta_c] = \frac{1}{\Lambda}. \quad (17)$$

We can figure out $E[\Delta_c]$ is $62.5\text{ns}$. Noticing that $E[\Delta_c]$ is much larger than the time duration of current receiving pulse, we can conclude intra-symbol interference primarily comes from the first cluster and the interference from other clusters can be neglected.

Assume that we are going to receive signals of the first ray from the first cluster. And correlation receiver is employed in the system, and the template for demodulation is

$$v(t) = \sum_{j=0}^{N_c-1} p(t - jT_f - C_j^{(1)}T_c - T_1 - \tau_{1,1}). \quad (18)$$

The output decision variables of correlation receiver are given by

$$Z = \int_{iT_f}^{(i+1)T_f} r(t)v(t)dt = Z_u + Z_n + Z_{IASI} + Z_{ISI} + Z_{MUI}. \quad (19)$$

where, $Z_u$, $Z_n$, $Z_{IASI}$, $Z_{ISI}$, $Z_{MUI}$ account for desired signal, additive white Gaussian noise, intra-symbol interference, inter-symbol interference and multiuser interference, respectively.

The energy for the desired signal $E_b$ and the energy for white Gaussian noise $\sigma_n^2$ can be expressed as

$$E_b = E(Z_u)^2 = F(\omega_0)\Omega_0 N_s^2.$$
$$\sigma_n^2 = E(Z_n)^2 = F(\omega_0)N_s N_0/2. \quad (20)$$

where, $\Omega_0$ is given by

$$\Omega_0 = \frac{1}{\gamma_0[(1-\beta)\lambda_1 + \beta\lambda_2 + 1]}. \quad (21)$$

And the variance of intra-symbol interference $\sigma_{IASI}^2$ is expressed as

$$\sigma_{IASI}^2 = E(Z_{IASI})^2$$
$$= F(\omega_0)E[\int_{iT_f}^{(i+1)T_f}[\sum_{l=1}^{L}\sum_{k=1}^{K}\alpha_{k,l}s^{(1,1)}(t-T_l-\tau_{k,l}) -$$
$$\alpha_{1,1}s^{(1,1)}(t-T_1-\tau_{1,1})]v(t)dt]^2$$
$$= F(\omega_0)N_s^2\sum_{l=1}^{L}\sum_{k=1}^{K}E[\alpha_{k,l}^2]E[R^2(\tilde{\tau}_{k,l}-\tilde{\tau}_{1,1})] +$$
$$F(\omega_0)N_s^2\sum_{l=1}^{L}\sum_{k=1}^{K}\sum_{l_1=1}^{L}\sum_{k_1=1}^{K}E[\alpha_{k,l}\alpha_{k_1,l_1}]E[R(\tilde{\tau}_{k,l}-\tilde{\tau}_{1,1})R(\tilde{\tau}_{k_1,l_1}-\tilde{\tau}_{1,1})].$$
$$(22)$$

where, $\tilde{\tau}_{k,l} = T_l + \tau_{k,l}$, and $R(\cdot)$ represents the autocorrelation function of the transmitted pulse waveform, and $k$ and $l$ cannot equal to 1 simultaneously. Moreover, $k_1$ and $k$, $l_1$ and $l$ should not be the same simultaneously.

Under these conditions, we can get $E[\alpha_{k,l}\alpha_{k_1,l_1}]$ is zero and the second term of the above equation is zero. Given that only the first cluster is taken into consideration and $\tilde{\tau}_{1,1}$ equals zero, the equation can be further simplified into

$$\sigma_{IASI}^2 = F(\omega_0)N_s^2\sum_{k=2}^{K}\int_0^{T_m}\Omega_0\exp(-y/\gamma)f_p(y)R^2(y)dy. \quad (23)$$

where, $y$ denotes $\tau_{k,l}$.

Since one information symbol usually comprises several pulses, for each received pulse, we can attribute the multipath interference from its previous pulses to inter-symbol interference (ISI). Thus, the number of interfering pulses is equivalent to $N_I N_s$. Meanwhile, Poisson process has the characteristic of memorylessness, so we can apply the similar method as IASI to analyze ISI and just change the energy of the first interfering ray $\Omega_0$ to $\Omega_\Sigma$. $\Omega_\Sigma$ is the sum of the average energy of the first ray of all the interfering pulses. Considering that different users adopt the homogeneous time hopping sequence, $\tau_{code(i)}^{(n)}$ has the same distribution for different $n$ and $i$. Therefore, we use $\tau_{code}$ to represent them all.

Hence, the variance for inter-symbol interference (ISI) is given by

$$\sigma_{ISI}^2 = E(Z_{ISI})^2 = F(\omega_0)N_s^2\sum_{k=2}^{K}\int_{-T_m}^{T_m}\Omega_\Sigma e^{-y/\gamma}f_p(y)R^2(y)dy.$$
$$(24)$$

where, $\Omega_\Sigma$ can be expressed as

$$\Omega_{\Sigma} = \sum_{s=1}^{N_l N_s -1} E[\Omega_s]$$

$$= \frac{1}{2T_s} \sum_{l=1}^{L} \sum_{s=1}^{N_l N_s -1} \int_{-T_s}^{T_s} \int_0^{sT_f + \tau_{code}} \int_{sT_f + \tau_{code}}^{\tau_{max}} \Omega_0 \, e^{-T_l/\Gamma} \, e^{-(sT_f + \tau_{code} - T_l)/\gamma}$$

$$\times f_c(T_l) f_c(T_{l+1}) dT_l dT_{l+1} d\tau_{code}. \quad (25)$$

where, $\Omega_s$ represents the $s$-th interfering pulse's energy.

The analysis of multiuser interference (MUI) is similar to IASI, and we just add $t_u$ to the whole delay. $t_u$ is the delay of other users relative to the first user. Suppose there exists $N_u + 1$ users in the system, the variance of MUI is

$$\sigma_{MUI}^2 = E(Z_{MUI})^2$$

$$= F(\omega_0) R_b N_s^2 N_u \sum_{k=2}^{K} \int_{-T_f/2}^{T_f/2} \int_{-z}^{T_m-z} \Omega_0 \, e^{-y/\gamma} f_p(y) R^2(y+z) dy dz$$

$$+ F(\omega_0) R_b N_s^2 N_u \sum_{k=2}^{K} \int_{-T_f/2}^{T_f/2} \int_{-z}^{T_m-z} \Omega_{\Sigma} \, e^{-y/\gamma} f_p(y) R^2(y+z) dy dz.$$

$$(26)$$

where, $y$ represents $\tau_{k,l}$ and $z$ represents $t_u$.

Finally, the signal to interference plus noise ratio (SINR) can be written as

$$\text{SINR} = \frac{E_b}{\sigma_n^2 + \sigma_{IASI}^2 + \sigma_{ISI}^2 + \sigma_{MUI}^2}. \quad (27)$$

And for BPSK system the bit error probability BER is

$$BER = \frac{1}{2} erfc(\sqrt{\frac{\text{SINR}}{2}}). \quad (28)$$

IV. SIMULATION RESULTS

The conventional second order derivation of Gaussian pulse waveform with time duration $T_m = 0.5$ns is adopted in simulation and its 10dB bandwidth is 5.6GHz. Without loss of generality, we consider $N_s$ equals 1. Plus, we set the hopping width $T_c$ equals $T_m$ and the number of hops $N_h$ is chosen to be 16. All the simulation is conducted under the indoor office LOS environment and data transmission rate is chosen to be 1Mbps and 15Mbps, respectively. The results are shown in Fig. 1.

As is shown in Fig. 1, the assumption that there does not exist intra-symbol interference (IASI) does not match with simulation, namely IASI cannot be neglected in the analysis. In addition, the figure reveals that IASI is a primary impact factor to system analysis because even for high $E_b/N_0$ values, the BER is still high. On the other hand, we can conclude that inter-symbol interference (ISI) has very little effect to the system performance when the data transmission rate is low (in the simulation we set 1Mbps and its simulation curve is very close to AWGN channel).

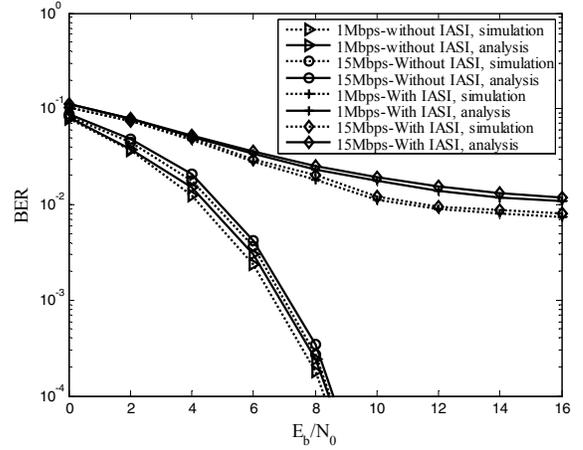

Figure 1. Analysis and simulation results with and without intra-symbol interference

Fig. 2 presents the BER performance of UWB with 1 user, 2 users, 4 users and 8 users, respectively. And each user's transmission data rate is set to be 15Mbps.

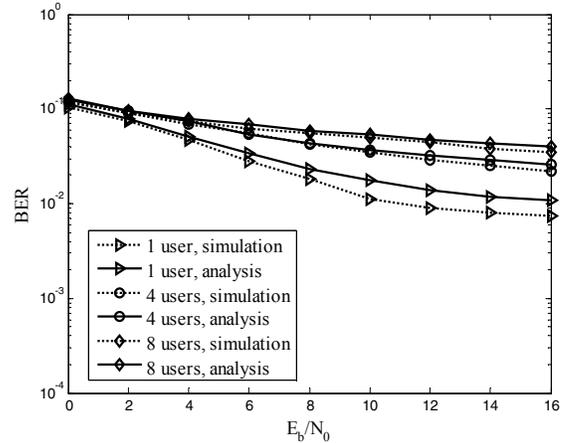

Figure2. Analysis and simulation results with multiuser interference

From Fig. 2, we can see that the analysis curves are very close to simulation curves, and the BER formulation have a very accurate evaluation of system performance in the presence of noise and different interferences. Furthermore, we can see that multiuser interference will further worsen system transmission performance compared with single user system.

Another issue is the discrepancy between analysis and simulation results. It should be noted that the simulation plots have a better BER performance than the analysis plots especially when $E_b/N_0$ is high. And there can be two reasons for the difference above: (1) In the analysis, we adopt the continuous second order derivation of Gaussian pulse with time duration from negative infinity to positive infinity to derive all the interference formulation, while in the simulation, a truncated waveform with time duration $T_m = 0.5$ns is employed. That is to say we introduce more interference in the analysis and as a result, a worse BER performance is shown in the analysis plots. (2) We adopt a simple way to describe the coming of random rays, as is talked in Section Ⅱ. Precisely speaking, the coming of rays should be modeled with a mixture of two Poisson processes with parameters ($\lambda_1, \lambda_2$). However, considering the low occurrence of Poisson process ($\lambda_1$) and also a simpler way of computation, the coming of rays is modeled as a single Poisson process ($\lambda_2$) in the analysis. Such simplification results in more interference in the analysis. It should be noted that the mean ray interval ($\Delta_\tau$) is in inverse proportion to $\lambda$ and $\lambda_2$ is far greater than $\lambda_1$: $\lambda_2 \gg \lambda_1$. In this case, we can conclude that the $\lambda_2$ Poisson process will have a much higher probability to generate a very dense multipath rays than the $\lambda_1$ Poisson process. Therefore, a single Poisson process (analysis results) means a more severe interference than a mixture of two Poisson processes (simulation results).

## V. CONCLUSION

By analyzing the indoor office LOS channel model defined by IEEE 802.15.4a Task Group, a system model for BER analysis of UWB systems with intra-symbol interference, inter-symbol interference, multiuser interference and AWGN is proposed. Furthermore, the variance for IASI, ISI and MUI is also derived as well as system BER formulation, and MATLAB simulation shows that the formulation can give an accurate evaluation of the system transmission performance. Moreover, this paper also proves that the intra-symbol interference should not be neglected by calculating the mean ray interval and comparing it with simulation. In the future, we will continue to analyze those different parameters in the system model, and try to improve system performance by parameter optimization.